\documentclass{elsart}
\usepackage{epsfig}
\begin{document}
\runauthor{Nishino, Okunishi, Hieida, Maeshima and Akutsu}
\begin{frontmatter}
\title{
Self-Consistent Tensor Product Variational Approximation for 3D Classical Models
}
\author[Kobe]{T.~Nishino}
\author[Suita]{K.~Okunishi}
\author[Gaku,Toyonaka]{Y.~Hieida}
\author[Toyonaka]{N.~Maeshima}
\author[Toyonaka]{Y.~Akutsu}
\address[Kobe]{Department of Physics, Faculty of Science, 
Kobe University, 657-8501, Japan}
\address[Suita]{Department of Applied Physics, Graduate school of Engineering,
Osaka University, Suita 575-0871, Japan}
\address[Gaku]{Computer Center, Gakushuin University,
Toshima-ku, Tokyo 171-8588, Japan}
\address[Toyonaka]{Department of Physics, Graduate School of Science, 
Osaka University, Toyonaka 560-0043, Japan}
\begin{abstract}
We propose a numerical variational method for three-dimensional (3D) classical
lattice models. We construct the variational state as a product of local tensors,
and improve it by use of the corner transfer matrix renormalization group
(CTMRG), which is a variant of the density matrix renormalization group (DMRG)
applied to 2D classical systems. Numerical efficiency of this
approximation is investigated through trial applications to the 3D Ising
model and the 3D 3-state Potts model. 
\end{abstract}
\end{frontmatter}

\section{Introduction}

The density matrix renormalization group (DMRG) \cite{DMRG1,DMRG2} has been
widely applied to one-dimensional (1D) quantum systems and two-dimensional
(2D) classical systems \cite{Dresden,Hallb}. A frontier in DMRG is to extend
its numerical algorithm to higher dimensional systems, chiefly for 2D quantum
and 3D classical systems. As far as the finite system algorithm is concerned,
decomposition of higher-dimensional clusters to 1D chains proposed by Liang and
Pang works efficiently \cite{zig}. On the other hand, we have not obtained any
satisfactory answer to extend DMRG toward infinite-size systems in higher
dimension. Nishino and Okunishi proposed a way of extending 
DMRG to 3D classical systems, which they call `the corner tensor renormalization
group (CTTRG)' \cite{CTTRG}, as a 3D generalization of both the transfer matrix
DMRG \cite{Dresden,Ni} and the corner transfer matrix renormalization group
(CTMRG) \cite{CTMRG1,CTMRG2} for 2D classical systems. Two major problems are
found in CTTRG when it is applied  to the 3D Ising model. One is that the
calculated transition temperature $T_{\rm c}$ is much higher than one of the 
most reliable $T_{\rm c}$ obtained by the Monte Carlo (MC) simulations
\cite{MC1,MC2}. The other problem is  the very slow decay of the the
density-matrix eigenvalues \cite{Peschel}, that spoils the numerical efficiency of
the block-spin transformation. 

\begin{figure}
\centerline{\epsfxsize=85mm \epsffile{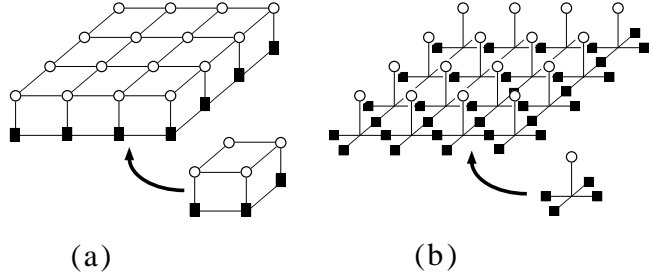}}
\caption{ Two representative constructions of the tensor product variational
state: (a) the IRF type and (b) the vertex type. The white circles denote 
spin (or field) variables, and the black squares denote auxiliary variables. We
consider the simplest example of the (a)-type product state in this paper. } 
\label{fig:1}
\end{figure}

A way of generalizing DMRG to higher dimensions is to investigate the variational
structure of DMRG, where the variational state for the transfer matrix or the
Hamiltonian is written in a product of orthogonal matrices \cite{Ostlund}.  Two
types of 2D tensor product states have been considered as higher-dimensional
extensions of this matrix product state. One is `the interaction round a face (IRF)'
type product states shown in figure (\ref{fig:1}(a)) \cite{Sierra}, and the other is
the vertex type one in figure (\ref{fig:1}(b)) \cite{Zitt2,Hieida}. For a 2D tensor
product state $V$, the variational energy and the variational partition function,
respectively, is written as 
\begin{equation}
\lambda = \frac{\langle V | \, H \, | V \rangle}{\langle V | V \rangle}
\,\,\,\, {\rm and} \,\,\,\,
\frac{\langle V | \, T \, | V \rangle}{\langle V | V \rangle} \, ,
\label{varf}
\end{equation}
where $H$ and $T$ denotes a Hamiltonian for a 2D quantum system and a transfer
matrix for a 3D classical system. Let us call such a variational estimation as `the
tensor product variational approximation (TPVA)' in the following. Calculation of
$\lambda$ have been performed by  MC simulation \cite{Zitt2}, product wave
function renormalization group (PWFRG) \cite{Hieida} or CTMRG \cite{Krm2}. A key
point  in TPVA is to find a good variational state $V$. So far, they assumed a
specific form of $V$, which contains several variational parameters, and tried to
find out the best $V$ by way of the parameter sweep. For example, Okunishi and
Nishino \cite{Krm2} investigated TPVA for the 3D Ising model, assuming $V$ in
the form of the Kramers-Wannier (KW) approximation \cite{Krm1}. Their
variational state contains two adjustable parameters, and the best $V$ is
obtained through a two-parameter sweep. Such an  intuitive construction of $V$
is, however, not always applicable; for example, we don't know how to extend the
KW approximation for the 3D Potts models \cite{Potts}. How can we obtain the best
$V$ automatically for higher-dimensional systems? We find an answer
for 3D classical systems.

In this paper we propose a self-consistent improvement for the tensor product
state $V$ by way of CTMRG. We choose the 3D Ising model as an example of the 3D
classical systems, and formulate our self-consistent method in terms of the Ising
model.  In the next section, we introduce the simplest 2D tensor product state, and
give the formal expression of the variational partition function $\lambda$ in
eq.(\ref{varf}). We then obtain the  self-consistent equation for $V$ in \S 3,
considering the variation $\delta \lambda / \delta V$. In \S 4 we propose a
numerical algorithm to solve the self-consistent equation. In \S 5 we
check the numerical efficiency and stability of this algorithm when it is applied
to the 3D Ising model and the 3-state ($q=3$) 3D Potts model. Conclusions are
summarized in \S 6.

\section{Tensor Product Variational State}

\begin{figure}
\centerline{\epsfxsize=72mm \epsffile{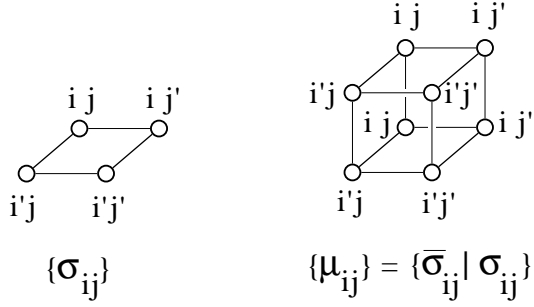}}
\caption{ Positions of the spin variables. The plaquett spin $\{ \sigma_{ij}^{~} \}$
in eq.(\ref{pspin}) consists of 4 neighboring spins, where  $i' = i+1$ and $j' =
j+1$, and the cube spin $\{ \mu_{ij}^{~} \}$  in eq.(\ref{pspin2}) consists of a
stack of two plaquett spins $\{ \bar\sigma_{ij}^{~} \}$ and  $\{ \sigma_{ij}^{~}
\}$. }  \label{fig:2}
\end{figure}

We briefly review the variational formulation of TPVA that was used for the
KW approximation of the 3D Ising model \cite{Krm2}. Let us
consider the 3D Ising model on the simple cubic lattice of the size $2N \times 2N
\times \infty$ to $X$, $Y$ and $Z$ directions, respectively, where open
(or fixed) boundary conditions are assumed for both $X$ and $Y$ directions. We are
interested in the bulk property of this model, and therefore suppose that the
system size $2N$ is sufficiently large. Suppose that the neighboring Ising spins
$\sigma$ and $\sigma'_{~}$ have ferromagnetic interaction $- J \sigma
\sigma'_{~}$. The transfer matrix $T$ from a $2N \times 2N$ spin layer
\begin{equation}
[ \sigma ] \equiv \left[ \begin{array}{ccccc}
\sigma_{1\,\,\,1} & \ldots & \sigma_{1\,\,N} & \ldots & \sigma_{1\,\,\,2N} \\
\vdots & \ddots &  \vdots & \ddots & \vdots \\
\sigma_{N\,\,1} & \ldots & \sigma_{N\,\,N}  & \ldots & \sigma_{N\,2N} \\
\vdots & \ddots & \vdots & \ddots & \vdots \\
\sigma_{2N\,1} & \ldots & \sigma_{2N\,N} & \ldots & \sigma_{2N\,2N} 
\end{array}
\right]
\label{lspin}
\end{equation}
to the next layer $[ \bar\sigma ]$ is then expressed as a product of local factors
\begin{equation}
T[ \bar\sigma  |  \sigma ] 
= \prod_{i=1}^{~2N-1~} \prod_{j=1}^{~2N-1~} X_{ij}^{~}
\equiv \prod_{ij} X_{ij}^{~} \, ,
\label{Trm}
\end{equation}
where $X_{ij}^{~}$ represents the Boltzmann factor for a local cube
\begin{eqnarray}
X_{ij}^{~} = {\rm exp} \biggl\{ \frac{K}{4} \!\!
&(& \!
\bar\sigma_{i'\!j} \bar\sigma_{i\,j} \! + \!
\bar\sigma_{i\,j'} \bar\sigma_{i\,j} \! + \!
\bar\sigma_{i'\!j'} \bar\sigma_{i'\!j} \! + \!
\bar\sigma_{i'\!j'} \bar\sigma_{i\,j'}
\biggr. \\
\biggl.
&+& \!
\bar\sigma_{i\,j} \sigma_{i\,j}^{~} \! + \!
\bar\sigma_{i'\!j} \sigma_{i'\!j}^{~} \! + \!
\bar\sigma_{i\,j'} \sigma_{i\,j'}^{~} \! + \!
\bar\sigma_{i'\!j'} \sigma_{i'\!j'}^{~}
\biggr. \nonumber\\
\biggl.
&+& \!
\sigma_{i'\!j}^{~} \sigma_{i\,j}^{~} \! + \!
\sigma_{i\,j'}^{~} \sigma_{i\,j}^{~} \! + \!
\sigma_{i'\!j'}^{~} \sigma_{i'\!j}^{~} \! + \!
\sigma_{i'\!j'}^{~} \sigma_{i\,j'}^{~} )
\biggr\} \nonumber
\label{Cube}
\end{eqnarray}
parameterized by $K = J / k_{\rm B}^{~} T$. We have used the notation $i' = i +
1$ and $j' = j+1$. (See figure (\ref{fig:2}).) We define $T[ \bar\sigma  | \sigma ]$
so that it is symmetric, because the symmetry simplifies the following
formulation.

The variational state in TPVA is a uniform product of local
tensors. In this paper, we focus on the simplest construction of the
tensor product state 
\begin{equation}
V[ \sigma ] = \prod_{ij}^{~} W_{ij}^{~}
= \prod_{ij} W\left( \begin{array}{cc}
\sigma_{i\,j}^{~} & \sigma_{i\,j'}^{~} \\
\sigma_{i'\!j}^{~} & \sigma_{i'\!j'}^{~}
\end{array} \right) \, ,
\label{prodS}
\end{equation}
where the local tensor $W_{ij}^{~}$ does not contains auxiliary variables, which
are shown by black squares in figure (\ref{fig:1}(a)) \cite{Hieida,Sierra}; to include
the auxiliary variables is straightforward, but makes the following equations
rather lengthy. The tensor product state $V[ \sigma ]$ is uniform in the sense that 
$W_{ij}^{~}$ is position independent. The local tensor $W_{ij}^{~}$ has 16
parameters, but not all of them are physically independent \cite{Bax}. For the
book keeping, let us use the notation
\begin{equation}
\{ \sigma_{i\,j}^{~} \} \equiv
\left( \begin{array}{cc}
\sigma_{i\,j}^{~} & \sigma_{i\,j'}^{~} \\
\sigma_{i'\!j}^{~} & \sigma_{i'\!j'}^{~}
\end{array} \right) 
\label{pspin}
\end{equation}
for the  plaquett spin, and write the local tensor $W_{ij}^{~}$ 
simply as $W\{ \sigma_{i\,j}^{~} \}$. In the same manner, let us write $X_{ij}^{~}$
as $X\{ \bar\sigma_{ij} | \, \sigma_{ij}^{~} \}$. (See figure (\ref{fig:2}).) 

Using $T[ \bar\sigma  |  \sigma ]$ and $V[ \sigma ]$ thus defined, the variational
partition function per layer is expressed as 
\begin{eqnarray}
\lambda &=& 
\frac{\displaystyle  \sum_{[ \bar\sigma ] [ \sigma ]}^{~}
V[ \bar\sigma ] \, T[ \bar\sigma  |  \sigma ] \, V[ \sigma ]}{ \displaystyle
\sum_{[ \sigma ]}^{~} \left( V[ \sigma ] \right)^2 }
=
\frac{\displaystyle  \sum_{[ \bar\sigma ] [ \sigma ]}^{~} \prod_{ij}
W\{ \bar\sigma_{i\,j}^{~} \} \, X\{ \bar\sigma_{ij} | \, \sigma_{ij}^{~} \} \, 
W\{ \sigma_{i\,j}^{~} \}}{\displaystyle
\sum_{[ \sigma ]}^{~}  \prod_{ij} 
\left( W\{ \sigma_{i\,j}^{~} \} \right)^2 } \nonumber\\
&=&
\frac{\displaystyle  \sum_{[ \bar\sigma ] [ \sigma ]}^{~} \prod_{ij}
G^1_{~}\{ \bar\sigma_{ij} | \, \sigma_{ij}^{~} \} }{ \displaystyle
\sum_{[ \sigma ]}^{~}  \prod_{ij} 
G^0_{~}\{ \sigma_{ij}^{~} \}} \equiv  \frac{Z^1_{~}}{Z^0_{~}} \, ,
\label{varf2}
\end{eqnarray}
where we have defined $G^0_{~}$ and $G^1_{~}$ as
\begin{eqnarray}
G^0_{~}\{ \sigma_{ij}^{~} \} 
&=& 
\left( W\{ \sigma_{i\,j}^{~} \} \right)^2 \, , \nonumber\\
G^1_{~}\{ \bar\sigma_{ij} | \, \sigma_{ij}^{~} \} 
&=& 
W\{ \bar\sigma_{i\,j}^{~} \} \, 
X\{ \bar\sigma_{ij} | \, \sigma_{ij}^{~} \} \, 
W\{ \sigma_{i\,j}^{~} \}  \, .
\label{defG}
\end{eqnarray}
It should be noted that $Z^0_{~}$ is a partition function of an IRF model \cite{Bax}
on $2N \times 2N$ square, and $Z^1_{~}$ is that of a 2-layer lattice model of the
same size.

\section{Self-Consistent Relation for the variational state}

Now we explain the self-consistent equation for the variational state $V[
\sigma ]$, the equation
which is satisfied when  $\lambda$ in eq.(\ref{varf2}) is maximized. Let us
consider the variation of $\lambda$ with respect to the variations of local tensors
\begin{equation}
\frac{\delta \, \lambda}{\delta \, V} \equiv \sum_{ij}^{~}
 \frac{\delta \, \lambda}{\delta \, W_{ij}^{~} }\, 
\label{diff}
\end{equation}
under the condition that the system size $2N$ is sufficiently large and the
boundary effect is negligible. Then most of the terms in the r.h.s. are almost the
same, and it is sufficient to consider the variation of  $\lambda$ with respect to
the local change $W_{NN}^{~}$  $\rightarrow W_{NN}^{~} + \delta \, W_{NN}^{~}$ at
the center of the system, where  $W_{NN}^{~}$ represents the local tensor at the
center. (See eqs.(\ref{lspin}) and (\ref{prodS}).)

The variation $\delta \, \lambda / \delta \, W_{NN}^{~}$ can be explicitly 
written down by use of two matrices. One is the diagonal matrix
\begin{equation}
A\{ \sigma_{NN}^{~} \} = 
\sum_{[ \sigma ]'}^{~} \prod_{(ij)\neq(NN)}^{~} \!\!\!\!
G^0_{~}\{ \sigma_{ij}^{~} \} \, ,
\label{matA}
\end{equation}
where $\sum_{[ \sigma ]'}^{~}$ denotes spin configuration sum for all the spins in
the layer $[ \sigma ]$ except for the spin plaquett $\{ \sigma_{NN}^{~} \}$
at the center; we
interpret $A\{ \sigma_{NN}^{~} \}$ as a 16-dimensional matrix
$M\{ \bar\sigma_{NN}^{~} | \sigma_{NN}^{~} \}$ where 
$M\{ \sigma_{NN}^{~} | \sigma_{NN}^{~} \} = A\{ \sigma_{NN}^{~} \}$
and is zero when $\{ \bar\sigma_{NN}^{~} \} \neq \{ \sigma_{NN}^{~} \}$.
From the definition, $Z^0_{~}$ in eq.(\ref{varf2}) is equal to $\sum_{\{
\sigma_{NN}^{~} \}}^{~} G^0_{~}\{ \sigma_{NN}^{~} \}  A\{ \sigma_{NN}^{~}
\}$. The other matrix is 
\begin{equation}
B\{ \bar\sigma_{NN}^{~} | \sigma_{NN}^{~} \}
= X\{ \bar\sigma_{NN} | \sigma_{NN}^{~} \} \!\!
\sum_{[ \bar\sigma ]' [\sigma ]'}^{~} \prod_{(ij)\neq(NN)}^{~} \!\!\!\!
G^1_{~}\{ \bar\sigma_{ij} | \, \sigma_{ij}^{~} \} \, ,
\label{matB}
\end{equation}
which is related to $Z^1_{~}$ as $Z^1_{~} = \sum_{\{ \bar\sigma_{NN}^{~} \} 
\{ \sigma_{NN}^{~} \}}^{~} W_{~}\{ \bar\sigma_{NN}^{~} \} 
B\{ \bar\sigma_{NN}^{~} | \sigma_{NN}^{~} \} W\{ \sigma_{NN}^{~} \}$.
By use of $A\{ \sigma_{NN}^{~} \}$ and $B\{ \bar\sigma_{NN}^{~}
| \sigma_{NN}^{~} \}$ thus created, we can write down $\lambda$ as
\begin{equation}
\lambda = \frac{Z^1_{~}}{Z^0_{~}}
= \frac{\displaystyle \sum_{\{ \bar\sigma \} \{ \sigma \}}^{~}
W\{ \bar\sigma \} \, B\{ \bar\sigma | \sigma \} \, W\{ \sigma \} }{
\displaystyle
\sum_{\{ \sigma \}}^{~} W\{ \sigma \} \, A\{ \sigma \} \, W\{ \sigma \}} \, ,
\label{varf4}
\end{equation}
where we have dropped the subscripts from  $\{ \sigma_{NN}^{~} \}$ and 
$\{ \bar\sigma_{NN} \}$ for book keeping. The condition $\delta \, \lambda / \delta
\, W\{ \sigma_{NN}^{~} \}$ $= 0$ draws the eigenvalue problem
\begin{equation}
\sum_{\{ \sigma \}}^{~} 
\frac{1}{A\{ \bar\sigma \}} \, B\{ \bar\sigma | \sigma \} \, W\{ \sigma \}
= \lambda \, W\{ \bar\sigma \}
\label{selfc}
\end{equation}
between the matrix $A^{-1}_{~}B$ and $W$; here we regard $W\{ \sigma \}$ as a
16-dimensional vector. This is the self-consistent equation that an optimized
tensor product state $V[ \sigma ]$ should satisfy.

\section{Numerical Algorithm of the Self-Consistent TPVA}

To use the self-consistent relation eq.(\ref{selfc}), we have to obtain
$A\{ \sigma \}$ and $B\{ \bar\sigma | \sigma \}$ for very large $N$. Though it is
impossible to obtain $A\{ \sigma \}$ and  $B\{ \bar\sigma | \sigma \}$ exactly,
the CTMRG \cite{CTMRG1,CTMRG2} enables us to numerically obtain them very
accurately. Let us introduce a new notation
\begin{equation}
\mu_{ij}^{~} \equiv \left( \bar\sigma_{ij}, \sigma_{ij}^{~} \right) \, ,
\label{vspin}
\end{equation}
which groups a pair of adjacent spins $\bar\sigma_{ij}$ and $\sigma_{ij}^{~}$.
Using $\mu_{ij}^{~}$, we can rewrite the stack of two plaquett spins
$\{ \bar\sigma_{ij} | \sigma_{ij}^{~} \}$ as 
\begin{equation}
\{ \mu_{i\,j}^{~} \}  
=
\left( \begin{array}{cc}
\mu_{i\,j}^{~} & \mu_{i\,j'}^{~} \\
\mu_{i'\!j}^{~} & \mu_{i'\!j'}^{~}
\end{array} \right) \, ,
\label{pspin2}
\end{equation}
$X\{ \bar\sigma_{ij} | \, \sigma_{ij}^{~} \}$ as $X\{ \mu_{ij} \}$, and
$G^1_{~}\{ \bar\sigma_{ij} | \, \sigma_{ij}^{~} \}$ as $G^1_{~}\{ \mu_{ij} \}$.
(See figure (\ref{fig:2})) We drop the subscripts from $\{ \mu_{i\,j}^{~} \}$ to
write it as $\{ \mu \} $ when its position is apparent. 

\begin{figure}
\centerline{\epsfxsize=40mm \epsffile{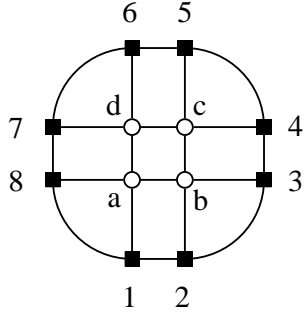}}
\caption{ Positions of the spin variables in eqs.(\ref{crctm}). The four white
circles denote the plaquett spin $\{ \sigma \}$ or $\{ \mu \}$ at the center, and
the black squares denote the block-spin variables used in CTMRG. } 
\label{fig:3}
\end{figure}

The matrices $A\{ \sigma \}$ and $B\{ \mu \} = B\{ \bar\sigma | \sigma \}$ can be
expressed as a combination of the corner transfer matrices (CTMs) and the
half-row transfer matrices (HRTMs), that appears when we apply CTMRG to both
the denominator and the numerator of eq.(\ref{varf2}) to obtain $Z^0_{~}$ and 
$Z^1_{~}$ \cite{Krm2}.
Let us write the CTM used for the calculation of $Z^0_{~}$ and $Z^1_{~}$,
respectively, as $C^0_{~}( \xi \sigma \xi'_{~} )$ and $C^1_{~}( \zeta \mu
\zeta'_{~} )$, where $\xi$, $\xi'_{~}$, $\zeta$, and $\zeta'_{~}$ are $m$-state
block spin variables. 
Also let us write HRTM as $P^0_{~}( \xi \sigma \sigma'_{~}
\xi'_{~} )$ and $P^1_{~}( \zeta \mu \mu'_{~} \zeta'_{~} )$ in the same manner.
Note that $C^0_{~}( \xi \sigma \xi'_{~} )$ and $P^0_{~}( \xi \sigma \sigma'_{~}
\xi'_{~} )$ are created from $G^0\{ \sigma \}$, and $C^1_{~}( \zeta \mu
\zeta'_{~} )$ and $P^1_{~}( \zeta \mu \mu'_{~} \zeta'_{~} )$ are from
$G^1\{ \mu \}$. Combining these CTMs and HRTMs,  $A\{ \sigma \}$ and 
$B\{ \mu \}$ are constructed as  
\begin{eqnarray}
\!\!\!\!\!\!A\{ \sigma \} =  \sum_{\xi_1 \ldots \xi_8}^{~} &&  
P^0_{~}( \xi_1^{~} \sigma_{a}^{~} \sigma^{~}_{b} \xi_2^{~} ) \, 
C^0_{~}( \xi_2^{~} \sigma_{b}^{~} \xi_3^{~} ) \,
P^0_{~}( \xi_3^{~} \sigma_{b}^{~} \sigma^{~}_{c} \xi_4^{~} ) \, 
C^0_{~}( \xi_4^{~} \sigma_{c}^{~} \xi_5^{~} ) \nonumber\\
&&
P^0_{~}( \xi_5^{~} \sigma_{c}^{~} \sigma^{~}_{d} \xi_6^{~} ) \,
C^0_{~}( \xi_6^{~} \sigma_{d}^{~} \xi_7^{~} ) \, 
P^0_{~}( \xi_7^{~} \sigma_{d}^{~} \sigma^{~}_{a} \xi_8^{~} ) \, 
C^0_{~}( \xi_8^{~} \sigma_{a}^{~} \xi_1^{~} )  \, , \nonumber\\
\!\!\!\!\!\!B\{ \mu \} =   \, X\{ \mu \} \! \sum_{\zeta_1 \ldots \zeta_8}^{~} &&
P^1_{~}( \zeta_1^{~} \mu_{a}^{~} \mu^{~}_{b} \zeta_2^{~} ) \, 
C^1_{~}( \zeta_2^{~} \mu_{b}^{~} \zeta_3^{~} ) \, 
P^1_{~}( \zeta_3^{~} \mu_{b}^{~} \mu^{~}_{c} \zeta_4^{~} ) \, 
C^1_{~}( \zeta_4^{~} \mu_{c}^{~} \zeta_5^{~} ) \nonumber\\
&&
P^1_{~}( \zeta_5^{~} \mu_{c}^{~} \mu^{~}_{d} \zeta_6^{~} ) \,
C^1_{~}( \zeta_6^{~} \mu_{d}^{~} \zeta_7^{~} ) \, 
P^1_{~}( \zeta_7^{~} \mu_{d}^{~} \mu^{~}_{a} \zeta_8^{~} ) \, 
C^1_{~}( \zeta_8^{~} \mu_{a}^{~} \zeta_1^{~} ) 
\label{crctm}
\end{eqnarray}
where the positions of the spin variables are shown in figure (\ref{fig:3}). In
principle, we can use $A\{ \sigma \}$ and $B\{ \mu \}$ thus constructed to solve
the self-consistent eq.(\ref{selfc}). 

\begin{figure}
\centerline{\epsfxsize=55mm \epsffile{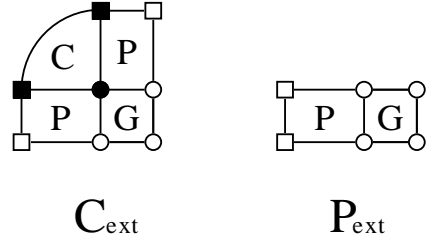}}
\caption{ Extension of CTM and HRTM \cite{CTMRG1,CTMRG2}. The local factor $G$
is created from the improved local tensor in eq.(\ref{improv}).
} 
\label{fig:4}
\end{figure}

To make the self-consistent improvement for $W\{ \sigma \}$ more efficiently,
we employ a numerical algorithm that simultaneously performs the extension of
the system size  in  CTMRG and the self-consistent improvement by
eq.(\ref{selfc}). The numerical procedures are as follows:
\vskip 10pt
\begin{itemize}
\item[(a)] Create $G^0_{~}\{ \sigma \}$ and $G^1_{~}\{ \mu \}$  from 
$X\{ \mu \}$ defined in eq.(\ref{Cube}) and the initial $W\{ \sigma \}$: 
\begin{eqnarray}
G^0_{~}\{ \sigma \} 
&=& 
\left( W\{ \sigma \} \right)^2 \, ,
\label{creG} \\
G^1_{~}\{ \mu \} 
\equiv G^1_{~}\{ \bar\sigma | \sigma \} &=&
W\{ \bar\sigma \} \, X\{ \bar\sigma | \, \sigma \} \, W\{ \sigma \} \, .
\nonumber
\end{eqnarray}
The choice of the initial $W\{ \sigma \}$ is not so relevant, since it is improved
afterward.
\item[(b)] Create the initial $C^0_{~}( \xi \sigma \xi'_{~} )$ and the initial
$P^0_{~}( \xi \sigma \sigma'_{~} \xi'_{~} )$ from $G^0_{~}\{ \sigma \}$, following
the standard initialization procedure in  CTMRG \cite{CTMRG1,CTMRG2}. Also
create $C^1_{~}( \zeta \mu \zeta'_{~} )$ and  $P^1_{~}( \zeta \mu \mu'_{~}
\zeta'_{~} )$ from $G^1_{~}\{ \mu \}$ in the same way. 
\item[(c)] Obtain the matrices $A\{ \sigma \}$ and 
$B\{ \mu \} = B\{ \bar\sigma | \sigma \}$ using eqs.(\ref{crctm}).
\item[(d)] Improve $W\{ \sigma \}$ by multiplying $A^{-1}_{~} B$
\begin{equation}
W_{\rm new}^{~}\{ \bar\sigma \} =
\sum_{\{ \sigma \}}^{~} \frac{1}{A\{ \bar\sigma \}} \, 
B\{ \bar\sigma | \sigma \} \, W_{\rm old}^{~}\{ \sigma \} \, ,
\label{improv}
\end{equation}
and normalize $W_{\rm new}^{~}\{ \sigma \}$ so that
\begin{equation}
\sum_{\{ \sigma \}}^{~} 
\left( W_{\rm new}^{~}\{ \sigma \} \right)^2 = 1
\label{norm}
\end{equation}
is satisfied.
\item[(e)] Recreate $G^0_{~}\{ \sigma \}$ and $G^1_{~}\{ \mu \}$ by substituting
$W_{\rm new}^{~}\{ \sigma \}$ into eqs.(\ref{creG}).
\item[(f)] 
Extend $P^0_{~}$ and $P^1_{~}$ to obtain $P^0_{\rm ext}$ and $P^1_{\rm ext}$,
respectively, by joining the recreated $G^0_{~}$ and $G^1_{~}$ as shown in
figure (\ref{fig:4}); the numerical details are shown in ref. \cite{CTMRG1,CTMRG2}.
Also extend $C^0_{~}$ and $C^1_{~}$ to obtain $C^0_{\rm ext}$ and $C^1_{\rm ext}$.
\item[(g)] 
Create density matrices from the extended CTMs, and diagonalizing them to
obtain the RG transformations $\xi_{\rm old}^{~} \sigma \rightarrow \xi_{\rm
new}^{~}$ and $\zeta_{\rm old}^{~} \mu \rightarrow \zeta_{\rm new}^{~}$,
where $\xi$ and $\zeta$ are $m$-state block spins. Then apply
the RG transformations to $P^0_{\rm ext}$, $P^1_{\rm ext}$
$C^0_{\rm ext}$ and $C^1_{\rm ext}$.
\item[(h)] 
Goto (c), and repeat (c)-(g) to improve $W\{ \sigma \}$
iteratively, and stop when $W\{ \sigma \}$ reaches its fixed point.
\end{itemize}
To summarize, we put three additional steps (a), (c) and (d) to the standard CTMRG
algorithm.

\section{Numerical Results}

\begin{figure}
\centerline{\epsfxsize=85mm \epsffile{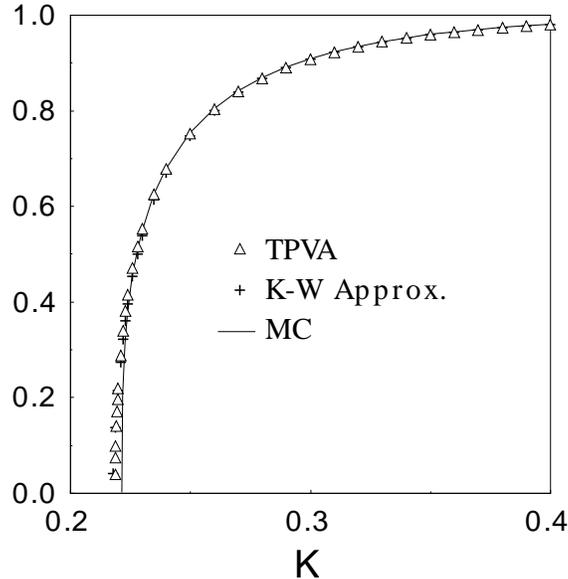}}
\caption{ Calculated spontaneous magnetization of the 3D Ising model. } 
\label{fig:5}
\end{figure}

Let us check the numerical efficiency and stability of the self-consistent
TPVA through trial applications to the 3D Ising model and the ferromagnetic $q=3$
Potts model. Figure (\ref{fig:5}) shows the  spontaneous magnetization $\langle
\sigma \rangle$ at the center of the $2N \times 2N \times \infty$ system, where
the curve, cross marks, and triangles, respectively, represent the result of the MC
simulation by Tarpov and Bl\"ote \cite{MC1}, KW approximation \cite{Krm2}, and
the self-consistent TPVA. We calculate  $\langle \sigma \rangle$ after repeating
the iteration (c)-(g) in the last section for $N = 10000$ times at most, keeping $m
= 10$ to $m = 20$ states for the block spin variables; the convergence with respect
to $m$ is very fast, where we obtain almost the same $\langle \sigma \rangle$ for
the cases $m = 10$ and $20$. The self-consistent improvement by
eq.(\ref{improv}) is monotonous in the whole parameter range, and no oscillatory
instability is observed. The calculated transition point $K_{\rm c} = 0.2188$ is 
about 1.3\% smaller than the MC result $K_{\rm c}^{\rm MC} = 0.2216544$. 

It turns out that the spontaneous magnetization calculated by KW approximation,
which gives the transition point $K_{\rm c}^{\rm KW} = 0.2180$, is quite close to
the result of the self-consistent (SC) TPVA. This means that the intuitive choice
of the variational state in KW approximation is actually very good, within the
simplest product state defined in eq.(\ref{prodS}). Inclusion of auxiliary
variables to the tensor product state is necessary for the further improvement
of the tensor product variational state. Note that the computational time required
for the KW approximation is several times larger than the self-consistent TPVA,
because the former finds the partition function extremum via 2-parameter sweep.

\begin{figure}
\centerline{\epsfxsize=95mm \epsffile{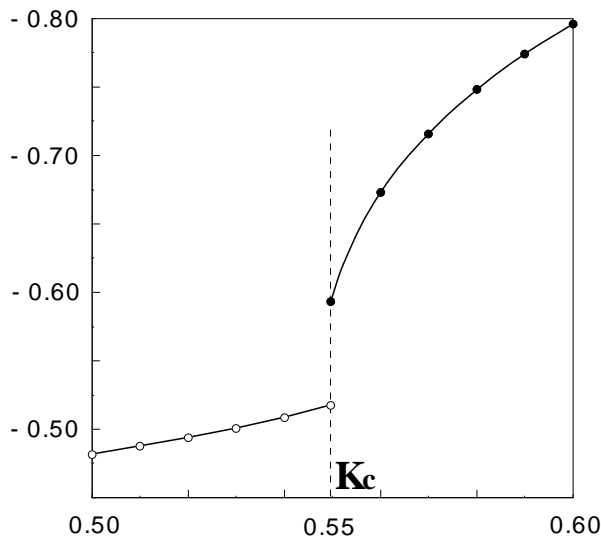}}
\caption{ Energy per bond $\langle -\delta( \sigma, \sigma'_{~}) \rangle$
of the ferromagnetic 3D $q=3$ Potts Model. } 
\label{fig:6}
\end{figure}

Figure (\ref{fig:6}) shows the energy per bond $E = \langle -\delta( \sigma,
\sigma'_{~}) \rangle$ of the ferromagnetic 3D $q=3$ Potts model, which is
calculated  by TPVA keeping  $m$ up to 15. The self-consistent improvement by 
eq.(\ref{improv}) is again monotonous, and $N = 1000$ is sufficient to get the
converged data; we need smaller $N$ for the Potts model than Ising model, because
the phase transition of the Potts model is first order. The calculated energy per
bond jumps from $E^{+} = -0.5173$ to $E^{-} = -0.5933$ at the calculated transition
point $K_{\rm c} = 0.54956$, where the calculated free energy of the  disordered
phase coincides with that of the ordered phase. The calculated transition point is
about 0.18\% smaller than one of the most reliable MC result $K_{\rm c}^{\rm MC}
= 0.550565 \pm 0.000010$ \cite{MC3}. The latent heat $l = 3(E^{+} - E^{-}) =
0.22769$  is about 41\% larger than the MC result $l = 0.16160 \pm 0.00047$
\cite{MC3}.

\section{Conclusion}

We have proposed a self-consistent TPVA, which gives the optimized tensor
product state for 3D classical systems, by way of the self-consistent
improvement of the local tensors. Since the method finds out the best variational
state without using a priori knowledge of the system, the self-consistent TPVA is
applicable for various 3D models described by  short range interactions. 

To generalize the self-consistent TPVA to 2D quantum systems is a next subject
that one might consider. This generalization  is not trivial, since we have used the
specific property of 3D classical systems when we obtain the self-consistent
equation. 

\ack

T.~N. thank to G.~Sierra and M.A.~Mart\'{\i}n-Delgado for the discussion about the
tensor product state at CSIC. K.~O. is supported by JSPS Research Fellowships for
Young Scientists.  This work was partially supported by the ``Research
for the Future'' Program from The  Japan Society for the Promotion of Science
(JSPS-RFTF97P00201) and by the Grant-in-Aid for  Scientific Research from
Ministry of Education, Science, Sports and Culture (No.~09640462 and
No.~11640376). Most of the numerical calculations were done by Compaq Fortran
on the HPC Alpha21264 Linux workstation.

\end{document}